# Small Distortions, Big Polarization: Tetragonal $BaTiO_3$ Nanoparticles for High-Performance Piezoelectric Nanogenerators

Shivshankar Jokare,[a,b] Vikash Kushwaha,[d] Mandar Shirolkar,[e,f] Shruti Kharadkar,[b] R. Boomishankar,[d] and Smita Chaturvedi* [a,b,c]

[a] *Department of Physics, Savitribai Phule Pune University, Pune, MH, India-411007.*
*Corresponding Author Email: smita.chaturvedi24@gmail.com*
[b] *Interdisciplinary School of Science, Savitribai Phule Pune University, Pune, MH, India-411007.*
[c] *Department of Physics, Indian Institute of Science Education and Research, Pune, MH, India-411008.*
[d] *Department of Chemistry, Indian Institute of Science Education and Research, Pune, MH, India-411008.*
[e] *Advanced Bio-Agro Tech Pvt. Ltd, Baner, Pune, 411045, Maharashtra, India*
[f] *Norel Nutrient Bio-Agro Tech Pvt. Ltd, Baner, Pune, 411045, Maharashtra, India*

Lead-free ferroelectric $BaTiO_3$ (BTO) nanoparticles with stabilized tetragonal distortion were synthesized via a ligand-assisted sol-gel method to serve as the active piezoelectric phase in high-performance hybrid piezoelectric nanogenerators (PENGs) operating with triboelectric synergy. XRD, Raman spectroscopy, and Pair Distribution Function (PDF) analysis confirmed pseudo-cubic tetragonal BTO nanoparticle, annealed at 900°C (BTO-900, ~54nm) and well-defined tetragonal BTO annealed at 1200°C (BTO-1200, ~104nm). Density Functional Theory (DFT) calculations confirmed enhanced Ti-O covalency, stronger Ti 3d-O 2p hybridization, narrower HOMO-LUMO gap, and improved dipole coherence in BTO-900, indicating superior ferroelectric polarization, piezoelectric activity, and charge retention. BTO-900-PDMS PENG device generated a high open-circuit voltage of ~100 V and stored ~29 µJ energy, and superior capacitor-charging performance of BTO-900, while the optimized BTO-1200 based device showed a higher power density of ~649 µW $cm^{-2}$ and short-circuit current of ~181 µA. The superior performance of BTO-900 was attributed to enhanced defect-assisted charge trapping, interfacial polarization, and impedance-controlled charge retention. By coupling the intrinsic piezoelectric response of covalency-engineered tetragonal BTO with triboelectric charge generation at the BTO-PDMS interface, this work establishes nanoparticle size and bonding engineering as an effective route to high-output, lead-free piezoelectric nanogenerators for self-powered electronics.

## 1. Introduction

The exponential growth of artificial intelligence (AI), machine learning (ML), cloud computing, and large-scale data centres has led to an unprecedented increase in global energy consumption. Modern AI-driven technologies, including autonomous systems, intelligent healthcare platforms, edge computing devices, and real-time data processing networks, require enormous computational power and continuous energy supply. In particular, the growing demand for high-performance computing and data storage infrastructure has significantly increased the energy footprint of data centres, raising concerns regarding sustainability, carbon emissions, and long-term energy management. Consequently, there is an urgent need for sustainable, maintenance-free, and self-powered energy solutions capable of supporting next-generation low-power electronic and distributed sensing systems.[1] Conventional electrochemical batteries remain the dominant power source for such systems; however, their limited lifetime, periodic replacement requirements, and environmental concerns associated with toxic waste and material extraction significantly restrict their long-term deployment in autonomous platforms.[2] Consequently, harvesting ambient mechanical energy from the surrounding environment has emerged as an attractive strategy for realizing self-powered electronics and sustainable sensing systems.

Among the available mechanical energy-harvesting technologies, piezoelectric nanogenerators (PENGs) have emerged as one of the most attractive approaches because they directly convert mechanical strain into electrical energy through the intrinsic piezoelectric polarization of the active material, offering high voltage output, a simple device architecture, lightweight construction, low fabrication cost, and excellent compatibility with flexible, self-powered platforms.[3,4] Since the pioneering demonstration of the piezoelectric nanogenerator by Wang and co-workers, PENGs and the closely related triboelectric nanogenerators (TENGs) have been widely explored for harvesting biomechanical energy, acoustic vibrations, fluid motion, and ambient mechanical stimuli.[5] The electrical output of such nanogenerators depends strongly on the piezoelectric polarization, surface charge density, and dielectric response of the active layer. In PENGs, embedding a high-permittivity ferroelectric ceramic into a flexible polymer matrix simultaneously boosts the piezoelectric potential and the dielectric charge-storage capability of the composite, making ceramic-in-polymer architectures an especially effective strategy for high-output, mechanically robust devices. In particular, coupling this intrinsic piezoelectric response with the triboelectric charging that develops at the polymer-electrode contact gives rise to hybrid piezoelectric-triboelectric nanogenerators, in which a dominant piezoelectric contribution is reinforced by triboelectric charge generation, producing significant improvements in voltage generation, current density, and power output.[6]

Barium titanate is one of the most extensively studied lead-free ferroelectric perovskites owing to its high dielectric constant, spontaneous polarization, strong piezoelectric response, and chemical stability.[7] Structurally, BTO crystallizes in the $ABO_3$ perovskite framework, where $Ba^{2+}$ occupies the A-site and $Ti^{4+}$ resides within an oxygen octahedral cage. The off-centre displacement of $Ti^{4+}$ ions inside the $TiO_6$ octahedra gives rise to spontaneous polarization and ferroelectricity. At room temperature, bulk BTO stabilizes in the tetragonal P4mm phase, which exhibits non-centrosymmetric distortion and strong piezoelectric activity.[8] Because of these properties, BTO has been widely investigated for applications in flexible electronics, actuators, non-volatile memories, dielectric capacitors, and energy-harvesting systems. Compared to toxic lead-based piezoelectric materials such as $Pb(Zr,Ti)O_3$ (PZT), BTO offers an environmentally benign alternative suitable for sustainable electronic technologies.[9]

BTO is known for its rich polymorphism and sequence of temperature-dependent phase transitions. In bulk form, BTO displays four temperature dependent crystallographic phases such as, cubic (*Pm*-3*m*, paraelectric), tetragonal (*P*4*mm*, ferroelectric), orthorhombic (*Amm*2, ferroelectric) and rhombohedral (*R*3*m*, ferroelectric). Phase transition occurs at defined temperatures such as cubic to tetragonal at general Curie temperature ($T_C$) ~393K, tetragonal to orthorhombic at ~278K and orthorhombic to rhombohedral at ~183K. Each crystal, below ~393K, exhibits non-centrosymmetric distortion and therefore results in ferroelectricity and spontaneous polarization.

At room temperature, bulk BTO stabilizes in tetragonal *P*4*mm* phase Ti off-centred displacement of $Ti^{+4}$ cation along the [001] direction, resulting in net polarization along the c-axis. Near Tc (393K), dielectric permittivity is in several thousand and further drops with increasing temperature as material transitions into cubic- paraelectric state.[8]

Synthesizing BTO with desired tetragonal structure in nanoscale, where ferroelectric properties are enhanced is still remains a formidable challenge. As particle size decreases below 20-100 nm, tetragonal distortion diminishes and paraelectric cubic-like phase predominates at room temperature leading to loss of dielectric and ferroelectric properties which often depends on synthesis conditions.[10] Several techniques have been developed for synthesis of BTO such as solid-state reaction, hydrothermal, solvothermal, molten salt methods, ball milling, sol-gel.[8] Among these, sol-gel approach is rapidly emerging as preferred method due to pure phase and high-quality synthesis of BTO, as this chemical process relies on hydrolysis and polycondensation of molecular precursors like metal alkoxides or acetates. This yields a highly homogenous gel which produces BTO upon calcination.[11]

In addition to crystal structure, recent studies suggest that electronic structure, bond covalency, oxygen vacancy concentration, and interfacial polarization significantly influence the electromechanical and triboelectric performance of BTO-based nanocomposites. Enhanced Ti-O covalency and localized electronic polarization can improve dipole alignment and charge retention, thereby promoting efficient mechanical-to-electrical energy conversion. Nevertheless, the correlation between synthesis induced structural evolution, electronic structure modulation, and piezoelectric-triboelectric output performance in nanoscale BTO systems remains insufficiently understood.

Herein, we report a systematic investigation of nanoscale tetragonal BTO synthesized via a ligand-controlled sol-gel approach, using different titanium precursors and calcination temperatures. The influence of precursor chemistry and acetylacetone-mediated hydrolysis control on phase purity, tetragonality, and defect evolution was comprehensively examined. Structural characterization using X-ray diffraction and Raman spectroscopy revealed the formation of pseudo-cubic tetragonal BTO at 900 °C and well-defined tetragonal BTO at 1200 °C. Density Functional Theory (DFT) calculations were further employed to investigate electron localization, Ti-O bond covalency, charge distribution, and ferroelectric stabilization mechanisms at the atomic scale. To evaluate their practical applicability, BTO nanoparticles were incorporated into a polydimethylsiloxane (PDMS) matrix to fabricate flexible hybrid piezoelectric nanogenerators (PENGs) operating with triboelectric synergy. The optimized BTO-PDMS PENG devices exhibited remarkable electrical output performance with high open-circuit voltage, enhanced current density, and significant power density under periodic mechanical tapping. Furthermore, the influence of calcination-induced structural and electronic modifications on charge transport, impedance behaviour, and capacitor charging capability was systematically correlated with nanogenerator performance.

## 2. Experimental Section

### 2.1 Materials

$Ba(CH_3COO)_2$ Barium acetate, ($Ti(OCH_2CH_2CH_2CH_3)_4$) titanium butoxide, $Ti(OCH(CH_3)_2)_4$, $CH_3COCH_2COCH_3$ acetylacetone (acac), $(CH_3)_2CHOH$ isopropanol (IPA) and $CH_3COOH$ glacial acetic acid were used as initial materials.

### 2.2. Synthesis of $BaTiO_3$

BTO nanoparticles (NPs) were synthesized via a sol-gel method using stoichiometric amounts of barium acetate and titanium butoxide in a 1:1 molar ratio. Initially, barium acetate was dissolved in a mixed solvent system containing glacial acetic acid and distilled water under continuous stirring. The solution was heated to approximately 70 °C to eliminate residual $BaCO_3$ impurities and obtain a clear precursor solution. In a separate vessel, titanium butoxide was dissolved in a mixture of acetylacetone (acac) and isopropanol (IPA) under constant stirring to form a stable titanium-acetylacetonate complex and thereby regulate the hydrolysis rate of the titanium precursor.

Subsequently, the transparent barium precursor solution was added dropwise to the titanium-acac solution under vigorous stirring to ensure homogeneous mixing and controlled sol formation. The resulting sol was heated at 90 °C to obtain gel, yielding a homogeneous gel network. The obtained gel was dried in an oven at 200 °C and subsequently calcined at 900 °C and 1200 °C for 4 h to obtain crystalline BTO nanoparticles. The samples calcined at 900 °C and 1200 °C were denoted as BTO-900 and BTO-1200, respectively.

### 2.3 Synthesis of PENG based on PDMS-$BaTiO_3$

#### a) PDMS-BTO elastomer film

Hybrid piezoelectric nanogenerator (PENG) devices were fabricated using polydimethylsiloxane (PDMS) incorporated with BTO nanoparticles (NPs). The PDMS elastomer matrix was prepared using the DOW Sylgard 184 silicone elastomer kit by mixing the base polymer and curing agent in a 10:1 weight ratio. Subsequently, BTO-900 and BTO-1200 nanoparticles were dispersed into the PDMS matrix at filler concentrations of 5, 10, 15, 20, and 25 wt% to investigate the effect of nanoparticle loading on device performance.

The homogeneous PDMS-BTO composite elastomers were subsequently cured at 60 °C for 3 h to obtain flexible composite films suitable for PENG device fabrication.

**b) PDMS-BTO based PENG device**

The fabricated PDMS-BTO composite film was sandwiched between two copper electrodes to assemble the hybrid piezoelectric nanogenerator (PENG) device. Copper wires were connected to the electrodes for electrical measurements, and the entire device was encapsulated using Kapton tape to provide electrical insulation as well as protection from external environmental effects. (SI-Figure 1)

### 2.4 Characterization techniques

The crystal structure and phase purity of the synthesized materials were investigated using X-ray diffraction (XRD) analysis performed on a Bruker D8 Advance diffractometer equipped with Cu-Kα radiation ($\lambda \approx 1.540$ Å). Rietveld refinement of the diffraction data was carried out using the FullProf Suite. Pair Distribution Function (PDF) experiment was performed on European Synchrotron Radiation Facility (ESRF), 71, Avenue des Martyrs, CS 40220, 38043 Grenoble Cedex 9, France.

Transmission electron microscopy (TEM) and high-resolution TEM (HRTEM) analysis of synthesized powder samples (drop-casted on a copper grid, after dispersing in ethanol) were carried out using JEOL JEM - ARM200F microscope equipped with Schottky field emission gun. The energy dispersive spectroscopy (EDS) analysis was performed using EDS module attached to the instrument. Raman spectroscopy was performed using a Renishaw in Via confocal Raman microscope operating in backscattering geometry with an $Ar^+$ laser excitation wavelength of ~532 nm.

The P-E hysteresis loop measurement was performed on the thin film of BTO-900 and BTO 1200 using a Sawyer-Tower circuit at room temperature. The thin film was prepared by drop-casting 20 mg of BTO dissolved in 0.5 ml of methanol onto the ITO-coated glass. The conductive Indium-gallium alloy was served as the electrodes. The ferroelectric measurements were performed on an aix ACCT TF2000E model hysteresis loop analyser. The hysteresis loop was obtained using the double-wave method by employing the positive up and negative down (PUND) function of the instrument at a 2.5 Hz frequency and 298 K. Leakage currents were dynamically measured at various voltage steps during the hysteresis loop measurements.

The electromechanical piezoelectric coefficient ($d_{33}$) of BTO-900 and BTO-1200 NPs were measured using a Piezo-Meter system (Piezotest). For the measurements, the synthesized nanoparticle powders were pelletized. Both BTO-900 and BTO-1200 powder samples were pressed into pellets with a diameter of 10 mm and a thickness of 0.2 mm. The BTO-900 and BTO-1200 pellets were then sintered at 1050 °C for 6h and 1200 °C for 8h, respectively. After sintering, the pellet's opposite surfaces were coated with silver (Ag) paint to form electrodes. The $d_{33}$ measurements were performed under a dynamic force of 0.25 N at 110 Hz.

### 2.5 Nanogenerator Measurement

For PENG measurements, composite films were fabricated by incorporating BTO-900 and BTO-1200 nanoparticles into a PDMS matrix at varying BTO nanoparticle amounts. Devices containing 5, 10, 15, 20 and 25 wt.% of each type of nanoparticle (BTO-900 and BTO-1200) were prepared to systematically evaluate the influence of sample loading. The energy-harvesting performance of the fabricated hybrid piezoelectric nanogenerator (PENG) devices was evaluated using a customized mechanical impact setup operating in periodic vertical compression mode. Electrical output measurements were carried out under a constant applied force of 21 N at a frequency of 8 Hz to investigate the device response under cyclic mechanical excitation.

## 3. Result and Discussion

### 3.1 Structural analysis

Rietveld refinement of the XRD patterns for BTO-900 and BTO-1200 confirmed the tetragonal phase of the synthesized BTO through the characteristic peak splitting near, along with additional splitting near 66 °and 75° as shown in Figure 1 (a, b). Figure 1(c-e) shows the deconvoluted (002)/(200), (202)/(220), and (103)/(301)/(310) reflections, further confirming tetragonal phase evolution.

BTO-900 crystalizes in pseudo-cubic tetragonal structure with lattice parameters a = b = 3.98 Å and c = 4.01 Å, with α = β = γ = 90°. While, BTO-1200 exhibited a well-defined tetragonal crystal structure with refined lattice parameters a = b = 3.99 Å and c = 4.03 Å, α = β = γ = 90°. The larger c/a distortion observed for BTO-1200 indicates enhanced tetragonality.The electron density (ED) distributions derived from the refined crystal structures are shown in Figure 2 (a, b) for BTO-900 and BTO-1200, respectively. ED was calculated using the reverse Fourier transformation of the structure factors obtained from Rietveld refinement according to:

$$\rho(x,y,z) = \frac{1}{V}\sum_{hkl} F_{hkl} \cdot e^{\{-2\pi i(hx+ky+lz)\}} \tag{1}$$

where ρ(x,y,z) represents the electron scattering density, $F_{hkl}$ denotes the structure factor, *h*, *k*, and *l* are the Miller indices, and V is the unit-cell volume.[12] The ED maps provide a real-space visualization of electron probability distribution and chemical bonding within the crystal lattice.

The ED surrounding Ba ions appears highly spherical and spatially separated from oxygen atoms, confirming the ionic nature of Ba-O interactions. In contrast, the Ti-O bonds exhibit pronounced bridge-like ED connecting the Ti centers with neighboring oxygen atoms, reflecting strong Ti-O covalency and significant orbital hybridization.[13] Compared to BTO-1200, BTO-900 sample displays more distinct electron-sharing contours between Ti and O atoms. In BTO-1200, the bridge-like electron density becomes comparatively weaker and more localized, indicating reduced Ti-O covalency and enhanced bond ionicity. This reduction in electron delocalization suggests that Ti 3d-O 2p hybridization is more effective in BTO-900 than in BTO-1200, which may strongly influence the polarization behavior of the material.[14]

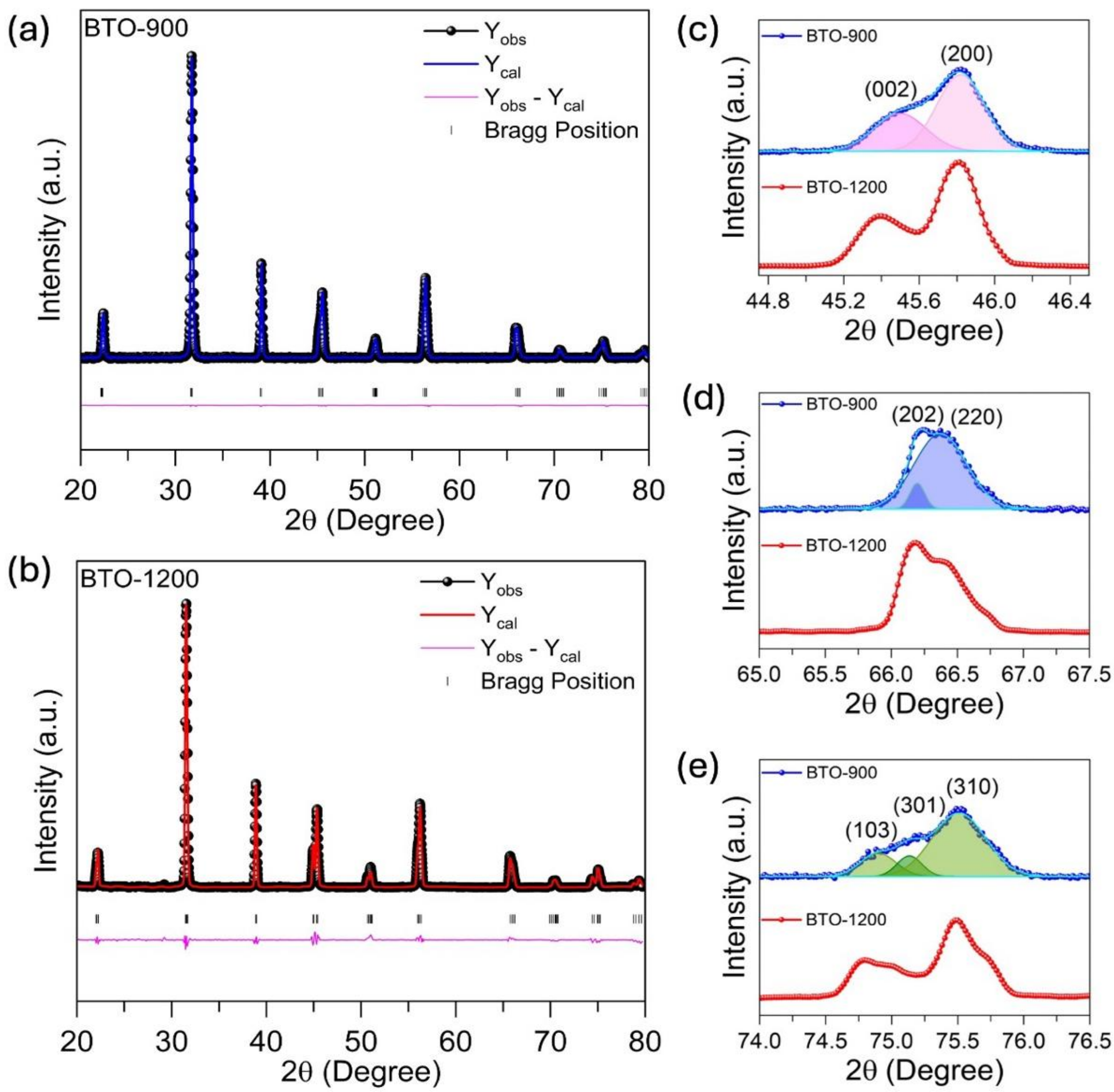

**Figure 1** (a) and (b) Rietveld refined XRD images of BTO-900 and BTO-1200 respectively. (c) Deconvoluted (002)/(200) reflections. (d) Deconvoluted (202)/(220) reflections. (e) Deconvoluted (103)/(301)/(310) reflections showing tetragonal phase evolution.

### 3.2 Pair Distribution Function Analysis

Conventional X-ray diffraction (XRD) primarily probes long-range crystallographic order through coherent scattering. However, in nanoscale BTO systems, local structural distortions are often obscured by micro strain effects, finite crystallite size, and domain broadening, causing the material to appear pseudo-cubic despite the presence of local ferroelectric distortions.[15] To overcome these limitations, Pair Distribution Function (PDF) analysis was employed, as it probes the real-space atomic arrangement and is highly sensitive to short-range and nearest-neighbor bonding environments.[16] The wide-range PDF refinement indicated distinct local ferroelectric distortions in both samples. Structural fitting using the *P4mm* framework confirmed the presence of local Ti off-centering within the $TiO_6$ octahedra, indicating the persistence of tetragonal ferroelectric distortion even at the nanoscale.[17] The

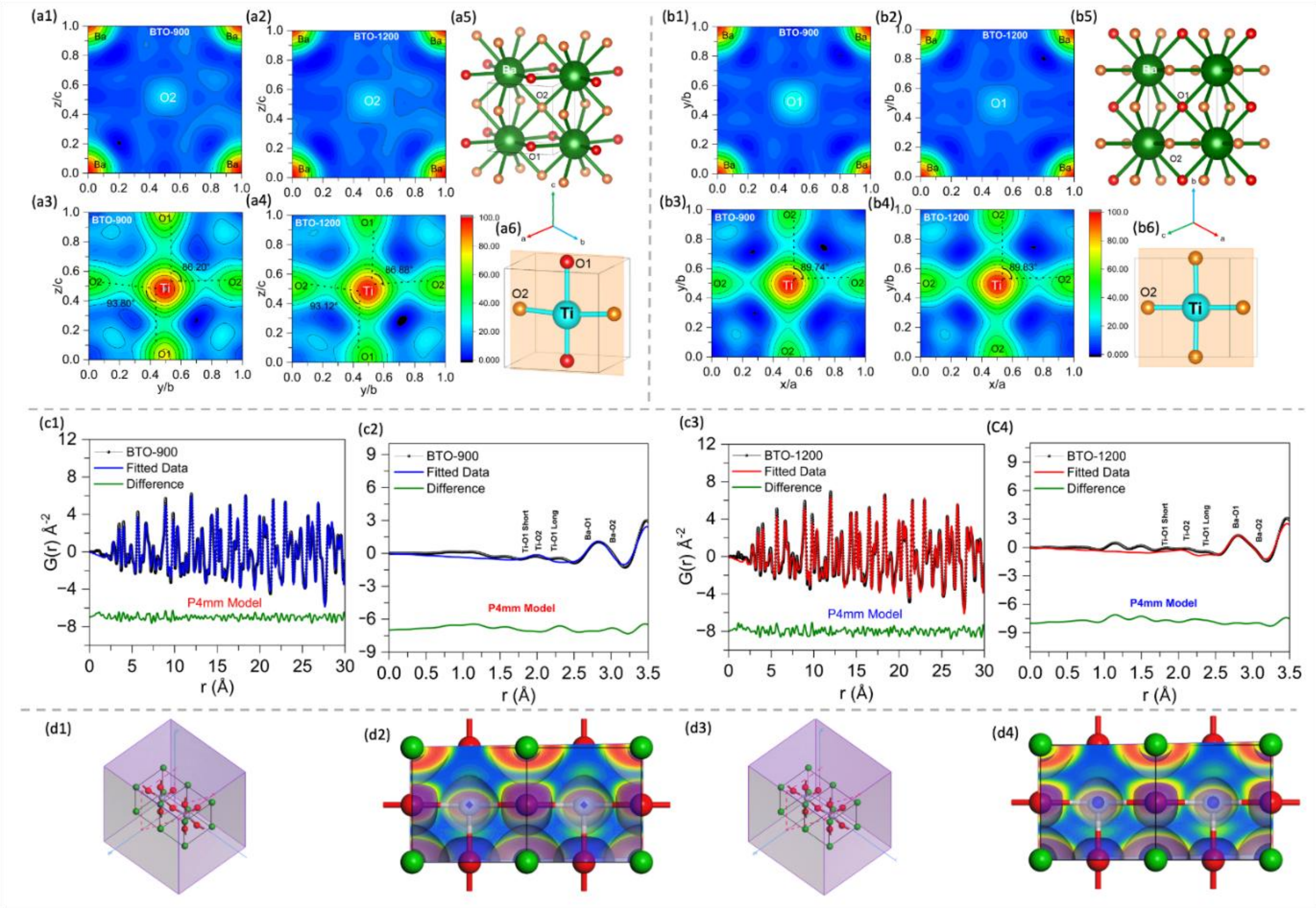

**Figure 2** (a1-a4) Electron density contour maps of BTO-900 and BTO-1200 showing electron distribution around Ba, Ti, O1, and O2 atoms with corresponding Ti-O bond angles. (a5, a6) Schematic tetragonal $BaTiO_3$ crystal structure and TiO6 octahedron. (b1-b4) Electron density contour maps along complementary crystallographic planes for BTO-900 and BTO-1200. (b5,b6) Structural models corresponding to the projected atomic arrangements. (c1,c3) Experimental and fitted PDF profiles of BTO-900 and BTO-1200 using the tetragonal *P*4*mm* model. (c2,c4) Enlarged PDF fitting profiles showing Ti-O and Ba-O correlations. (d1,d3) first Brillouin zone, and (d2, d4) electron localization function (ELF) contour maps representing orbital sharing among cations and anions for BTO-900 and BTO-1200 respectively.

fitted PDF profiles for BTO-900 and BTO-1200 are shown in Figure-2 (c1-c4). The refined structural parameters obtained from the PDF analysis are summarized in SI-Table 1.

The PDF and XRD results show that the refined lattice parameters and tetragonal characteristics are generally consistent, with only minor differences in the obtained values. The slight variation observed in the c lattice parameter is likely associated with microstrain effects present in the nanoscale samples. In addition to lattice parameters, the PDF analysis provides detailed information about local bond lengths involving oxygen atoms and their coordination with Ba and Ti ions. BTO-900 exhibits slightly higher average tetragonality compared to BTO-1200, which may be related to the transition from single-domain to multi-domain ferroelectric structures with increasing particle size. However, the real-space bond analysis obtained from PDF refinement reveals that BTO-1200 possesses a larger Ti-O1 bond length splitting ($\Delta$ = 0.270) compared to BTO-900 ($\Delta$ = 0.139), indicating stronger local ferroelectric distortion and enhanced local polarization in BTO-1200.

### 3.3 Electron Localization function (ELF) Distribution

DFT analysis (details of calculations in SI) was carried out to further validate experimentally observed bonding characteristics and their influence on the electronic structure of BTO. Figure 2 (d1, d3) represent first Brillouin zone for BTO-900 and BTO-1200 respectively, and Figure 2 (d2, d4) shows electron localization function (ELF) contour maps representing orbital sharing among cations and anions in BTO-900 and BTO-1200 respectively.

The Mulliken population analysis indicates important differences between BTO-900 and BTO-1200. The Ti-O bond population for BTO-900 (0.77) is slightly higher than that of BTO-1200 (0.75), indicating stronger orbital overlap and enhanced Ti-O covalency in BTO-900. Enhanced Ti-O covalency promotes stronger Ti 3d-O 2p hybridization, which in turn contributes to improved lattice stability and enhanced ferroelectric polarization. These calculations also point out differences in the O-O interactions. In BTO-1200, the O-O bond populations are negative (-0.07), suggesting repulsive electronic interactions within the lattice. In contrast, BTO-900 exhibits positive O-O bond populations (0.12 and 0.07), indicating more cooperative orbital interactions and improved lattice packing. This reduction in repulsive interactions is expected to enhance structural stability and dielectric behavior. The Hirshfeld charge analysis, further supports these observations. BTO-900 shows slightly lower positive charges on Ti and Ba ions (+0.848 and +1.342, respectively) compared to BTO-1200 (+0.850 and +1.341). This reduced ionicity suggests improved charge delocalization within the lattice, which can reduce internal polarization stress and favor long-range ferroelectric ordering. [18]

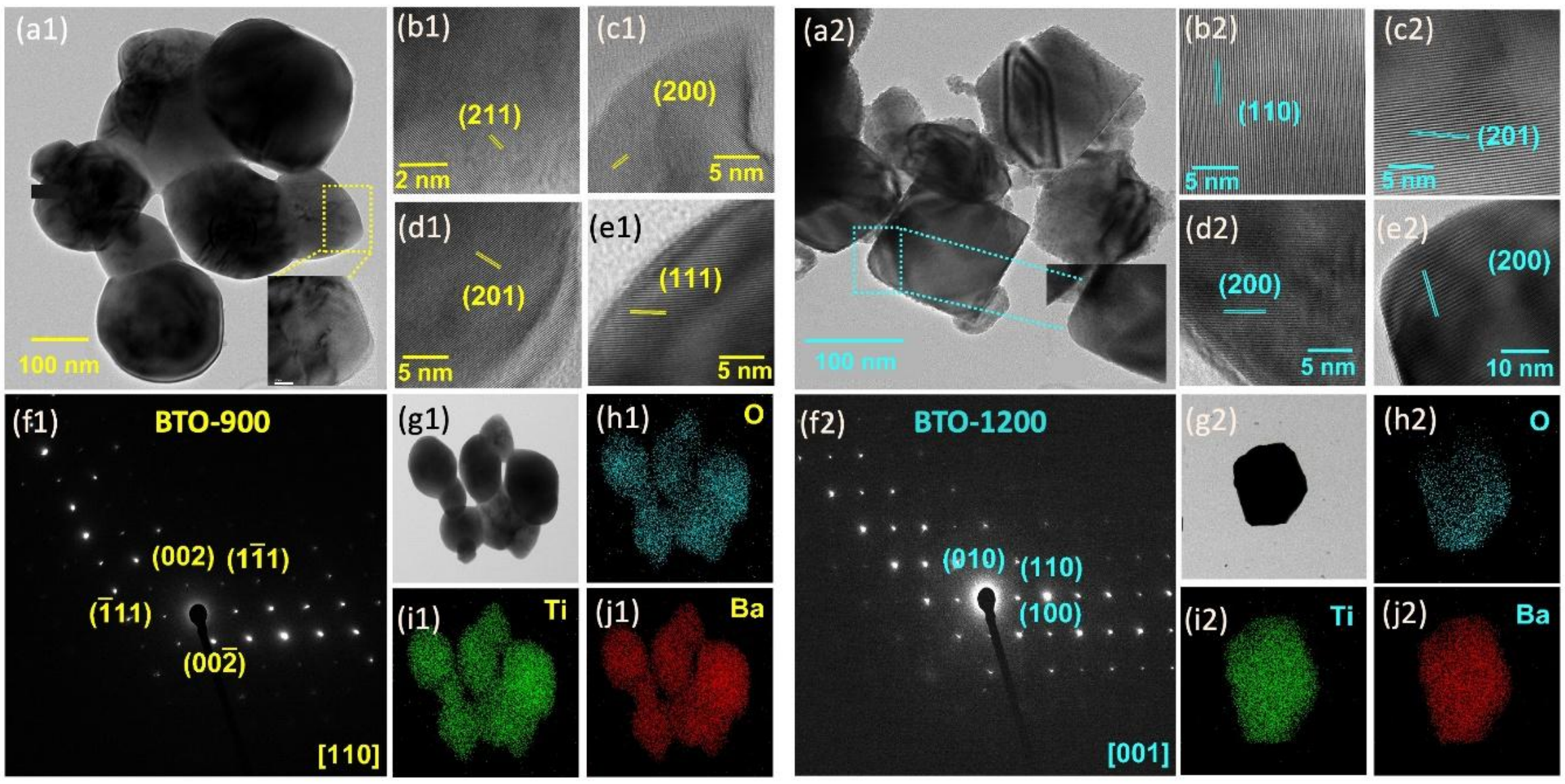

**Figure 3** (a1-e1) TEM and HRTEM images of BTO-900 showing lattice fringes indexed to the (211), (200), (201), and (111) planes. (f1) SAED pattern of BTO-900 indexed along the [110] zone axis. (g1-j1) STEM image and corresponding elemental mapping of O, Ti, and Ba for BTO-900. (a2-e2) TEM and HRTEM images of BTO-1200 showing well-defined lattice fringes corresponding to the (110), (201), and (200) planes. (f2) SAED pattern of BTO-1200 indexed along the [001] zone axis. (g2-j2) STEM image and elemental mapping of O, Ti, and Ba for BTO-1200, confirming uniform elemental distribution.

The electronic behaviour of a material can be understood from the energy difference between the Highest Occupied Molecular Orbital (HOMO) and Lowest Unoccupied Molecular Orbital (LUMO). In BTO-1200, stronger ionicity and more localized charge distribution result in a wider HOMO-LUMO gap, which can hinder ferroelectric switching behaviour. While BTO-900 exhibits stronger Ti-O covalency and improved O-O interactions, leading to a narrower HOMO-LUMO gap and enhanced electronic polarizability.[19] This favours easier dipole alignment and ferroelectric domain formation, indicating that BTO-900 possesses a more favourable electronic structure for ferroelectric behaviour.

Insight into the electronic structure was obtained from ELF maps, (Figure 2 (d2 and d4)), which provide a direct visualization of electron density localization within the crystal lattice. For BTO-900, the ELF maps reveal enhanced electron localization around the Ti-O bonds, indicating stronger covalent interactions and a more coherent charge distribution.[18] Such electronic characteristics support efficient dipole formation and improved ferroelectric domain alignment, which are essential for strong piezoelectric and ferroelectric responses. In contrast, the ELF maps of BTO-1200 exhibit comparatively diffuse electron density distribution, consistent with increased ionicity and weaker orbital overlap.[20]

Overall, based on above structural analyses, BTO-900 is superior in terms of ferroelectric traits to BTO-1200 for following reasons (i) enhanced Ti-O covalency, which gives stronger orbital overlap, narrower HOMO-LUMO gap and improved polarizability, (ii) reduced O-O repulsion gives cooperative lattice stabilization, favourable for domain formation, (iii) balanced charge distribution infers lower ionicity, better delocalization, reduced polarization stress, (iv) stronger dipole coherence gives improved ferroelectric switching. **SI-Table-2** represents Mulliken Atomic Populations summary of BTO-900 and BTO-1200. **SI-Table-3** represents Bond Populations summary of BTO-900 and BTO-1200.

### 3.4 Transmission electron Spectroscopy

The microstructural evolution and crystallographic features of the synthesized BTO nanoparticles were investigated using TEM, HRTEM, SAED, and STEM-EDS elemental mapping, as shown in Figure 3. The TEM images in Figure 3 (a1, a2) reveal agglomerated polyhedral nanoparticles with relatively uniform morphology. The BTO-1200 sample exhibits larger and more well-defined grains than BTO-900, indicating enhanced grain growth and densification at higher sintering temperature.

The HRTEM images shown in Figure 3(b1-e1) for BTO-900 display distinct lattice fringes indexed to the (211), (200), (201), and (111) planes, confirming the formation of crystalline tetragonal $BaTiO_3$. Similarly, the BTO-1200 sample in Figure 3(b2-e2) exhibits highly resolved lattice fringes corresponding to the (110), (201), and (200) planes. The clearer and more continuous fringes observed for BTO-1200 suggest improved crystallinity, lattice ordering, and reduced structural defects upon high-temperature sintering. The SAED patterns presented in Figure 3(f1,f2) further validate the crystalline nature of both samples. The diffraction spots indexed along the [110] and [001] zone axes confirm the tetragonal perovskite structure of $BaTiO_3$. Notably, the sharper and brighter diffraction spots observed for BTO-1200 indicate distinct cuboidal and rectangular shapes indicating the system has thermodynamically minimized its total surface energy.[21]

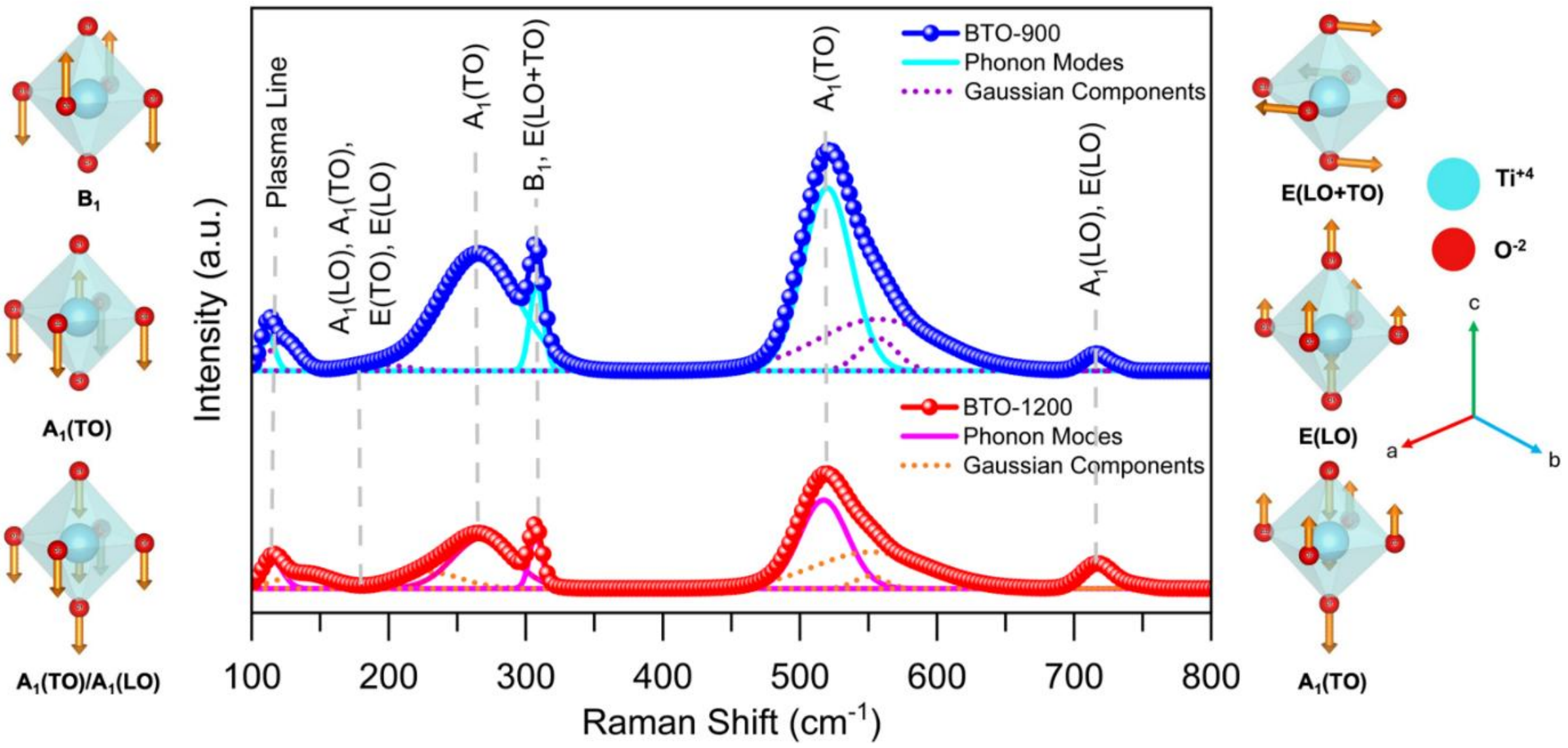

**Figure 4** Raman spectra and phonon mode analysis of BTO-900 and BTO-1200. Gaussian fitting components and phonon modes are shown along with schematic illustrations of the corresponding lattice vibrations of Ti and O ions within the $TiO_6$ octahedra.

The STEM image and corresponding elemental mapping results shown in Figure 3(g1-j2) demonstrate the homogeneous spatial distribution of Ba, Ti, and O elements throughout the particles for both samples. No noticeable elemental segregation or impurity-rich regions are observed, confirming the compositional uniformity and phase purity of the synthesized BTO nanoparticles. Overall, the microscopic analyses corroborate the XRD results and demonstrate that higher sintering temperature significantly enhances the crystallinity and structural ordering of the BTO.

### 3.5 Raman Spectroscopy

The room-temperature Raman spectra shown in Figure 4, for BTO-900 and BTO-1200 exhibit several characteristic vibrational modes confirming the non-centrosymmetric tetragonal structure of BTO. In ideal cubic BTO, the three $F_{1u}$ optical phonon modes are infrared-active and generally Raman inactive, while the $F_{2u}$ mode remains silent. However, in the tetragonal ferroelectric phase, Ti off-centering along the c-axis lowers the crystal symmetry, causing the splitting of the cubic $F_{1u}$ and $F_{2u}$ modes into Raman-active longitudinal optical (LO) and transverse optical (TO) phonons. Group theory predicts three $A_1$ modes, four doubly degenerate E modes, and one $B_1$ mode for tetragonal BTO with *P4mm* symmetry. The appearance of the $B_1$ mode near ~300 $cm^{-1}$ is particularly important, as this mode originates from the silent $F_{2u}$ mode of the cubic phase and serves as a signature of tetragonal distortion and ferroelectric symmetry breaking.[22]

**Table 1** Raman spectra summary for BTO-900 and BTO-1200

| Raman Mode | BTO-900 ($cm^{-1}$) | BTO-1200 ($cm^{-1}$) | Bulk ($cm^{-1}$)[23] | NP ($cm^{-1}$)[24] | Single Crystal ($cm^{-1}$)[25] | Thin Film ($cm^{-1}$)[26] |
|---|---|---|---|---|---|---|
| E(TO) | - | - | ~36 | | | |
| Plasma Line | ~117 | ~116 | | | | |
| $A_1$(TO) | | | ~170 | | ~173 | ~180 |
| E(TO), E(LO) | | ~180 | ~180 | | | |
| $A_1$(LO) | ~186 | | ~185 | | ~187 | |
| $A_1$(TO) | | | | | ~242 | |
| $A_1$(TO), E(TO) | ~265 | ~266 | ~270 | ~261 | | |
| $B_1$, E(LO+TO) | ~307 | ~307 | ~305 | ~304 | | |
| E(LO) | | | ~463 | | | |
| $A_1$(LO), E(TO) | | | ~486 | | ~485 | ~475 |
| E(TO) | ~519 | ~514 | ~518 | | | |
| $A_1$(TO) | | | ~520 | ~517 | ~522 | |
| $A_1$(TO) | | | | | | ~540 |
| $A_1$(LO), E(LO) | ~717 | ~716 | ~715-720 | ~715 | ~714 | |

DFT studies suggest that the ferroelectric distortion in BTO originates from Ti 3d-O 2p covalent hybridization. In the cubic phase, this hybridization contributes to transverse optical (TO) phonon vibrations involving the displacement of Ti ions against the surrounding oxygen octahedra. During the cubic-to-tetragonal phase transition, these phonon modes become frozen into the lattice, resulting in permanent Ti and O ion displacement characteristic of the ferroelectric P4mm phase.[27] The observed Raman bands were assigned to $A_1$(TO/LO), E(TO/LO), and $B_1$ vibrational modes of tetragonal BTO. The low-frequency $A_1$ modes associated with Ti-O vibrations along the polar c-axis appear in the ~180-270 $cm^{-1}$ region, while the mid-frequency $A_1$, E modes occur around ~500-520 $cm^{-1}$. A strong high-frequency $A_1$(LO) mode is observed near ~715-720 $cm^{-1}$, corresponding to Ti-O stretching vibrations along the polar axis.[22] The doubly degenerated E modes overlap with $A_1$ modes and contribute to the Raman bands near ~300-310 $cm^{-1}$ and ~512 $cm^{-1}$. These vibrational features are consistent with the expected phonon representation for tetragonal BTO:

$$3[A_1(TO) + A_1(LO)] + B_1 + 4[E(TO) + E(LO)] \quad (2)$$

The presence of well-defined $A_1$, E, and $B_1$ Raman modes at their characteristic frequencies confirms both the tetragonal crystal symmetry and ferroelectric nature of the synthesized BTO samples, particularly BTO-1200.[22]

Noticeably the Raman spectra of both BTO-900 and BTO-1200 does not have $A_1$(TO) around 175-185 $cm^{-1}$. This mode is associated with soft-phonon mode vibration. This is due to long-range Ti off-centring in ferroelectric BTO. Low-frequency soft-phonons are broadened and suppressed with decreasing particle size (due to finite-size effects and phonon confinement), while higher frequency bands are clearly observed.[28] So, absence of well-resolved ~175-185 $cm^{-1}$ Raman band $A_1(TO)$ (~180 $cm^{-1}$) is due to overdamping of soft-mode vibration which is arising from finite particle size and surface disorder, and does not suggest the loss of tetragonal symmetry. All observed modes and Raman shifts are summarised in **Table-1.**

### 3.6 Piezoelectric Coefficient ($d_{33}$) Measurement

BTO-900 exhibited a piezoelectric coefficient ($d_{33}$) of -26 pC/N, whereas BTO-1200 showed a $d_{33}$ value of 18 pC/N, without applying a conventional high-voltage poling process. The observed $d_{33}$ values confirm the intrinsic piezoelectric response arising from the non-centrosymmetric tetragonal crystal structure of BTO.
The difference in electromechanical response between the two samples is likely related to variations in particle size and defect concentration, both of which strongly influence ferroelectric domain behavior. The BTO-900 pellets, have relatively smaller grains and higher defect concentration, which can enhance domain-wall mobility and polarization response. While BTO-1200 with larger particles and comparatively reduced defect-assisted polarization effects.[29]

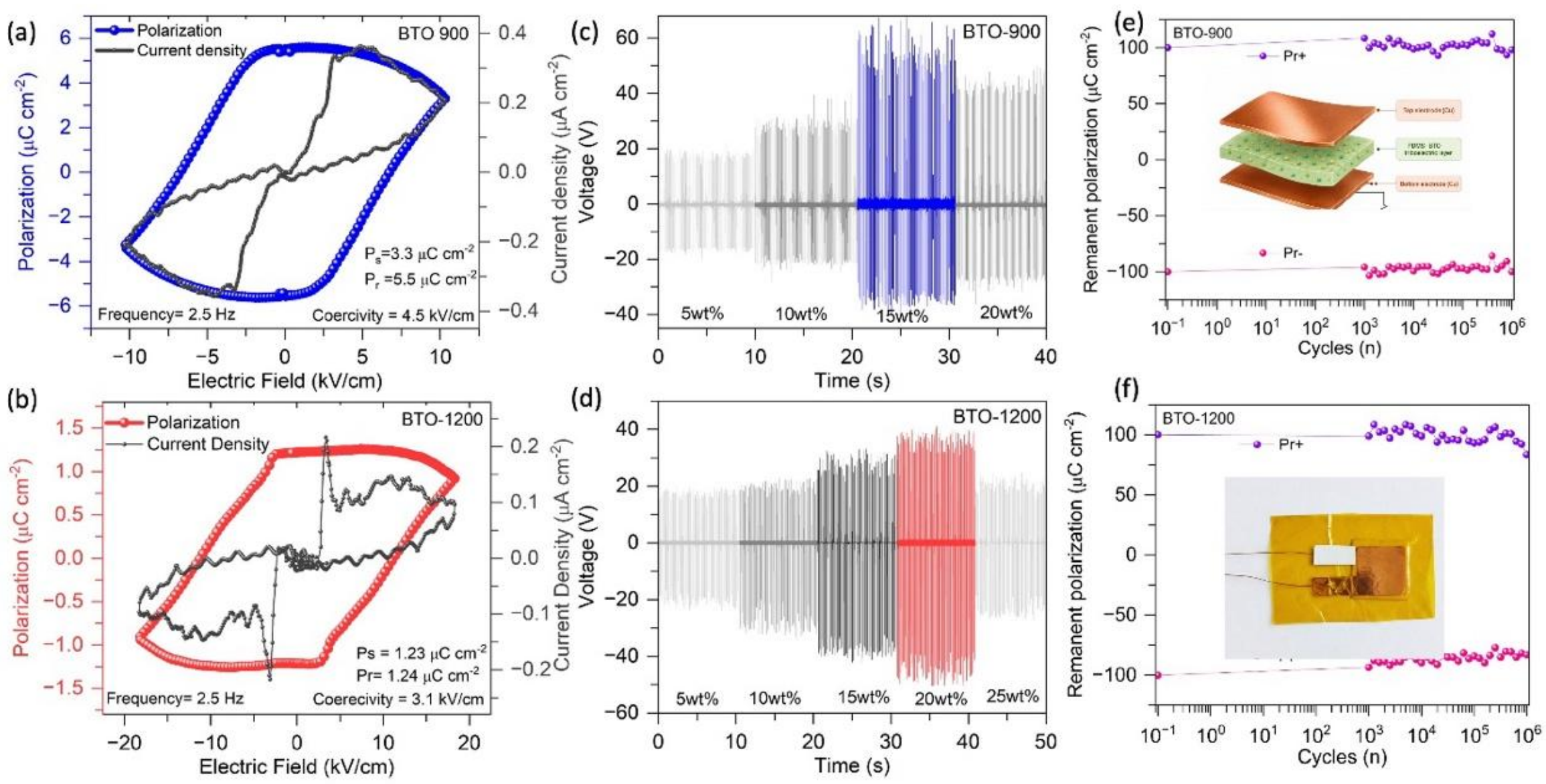

**Figure 5** (a,b) Polarization-electric field (P-E) hysteresis loops and current density curves of BTO-900 and BTO-1200 measured at 2.5 Hz. (c,d) Output voltage response of BTO-900 and BTO-1200 devices under different filler concentrations, demonstrating enhanced electrical output with optimized BTO loading. (e,f) Fatigue endurance behavior of BTO-900 and BTO-1200 up to cycles, showing stable remanent polarization. Insets show the schematic and fabricated flexible device structure.

### 3.7 Ferroelectric behaviour and nanogenerator performance

Polarization-electric field (P-E) hysteresis measurements were carried out for the BTO-900 and BTO-1200 films using the Positive-Up Negative-Down (PUND) method.[30] As shown in Figure 5(a,b), both samples exhibit well-defined rectangular ferroelectric hysteresis loops under an applied electric field of 10 kV cm at a frequency of 2.5 Hz, confirming their stable ferroelectric behavior.[31] The saturation polarization (Ps) values for BTO-900 and BTO-1200 were determined to be 3.3 C $cm^{-2}$ and 1.23 C $cm^{-2}$, respectively. Notably, BTO-900 exhibits significantly higher polarization (~2.75 times) than BTO-1200 under identical measurement conditions. The coercive fields of BTO-900 and BTO-1200 were found to be 4.5 and 3.1 kV $cm^{-1}$, respectively. The enhanced polarization observed for BTO-900 can be attributed to its smaller particle size and larger surface-to-volume ratio, which promote stronger electric-field interaction and polarization response. Additionally, Ti-O covalency in BTO-900 may further contribute to the improved ferroelectric characteristics.[32]

The current density-electric field (J-E) characteristics reveal comparatively low leakage current densities for both samples, remaining within the range of 0.4 A $cm^{-2}$. Distinct current-switching peaks were observed near the coercive field region for both BTO-900 and BTO-1200, corresponding to ferroelectric domain switching processes. The fatigue endurance measurements presented in Figure 5(e,f) demonstrate stable remanent polarization up to cycles, confirming the excellent cyclic stability and reliability of both BTO-based ferroelectric devices. [33]

The energy-harvesting performance of BTO-900-PDMS and BTO-1200-PDMS hybrid PENG devices was evaluated under periodic mechanical compression (21N, 8Hz). As

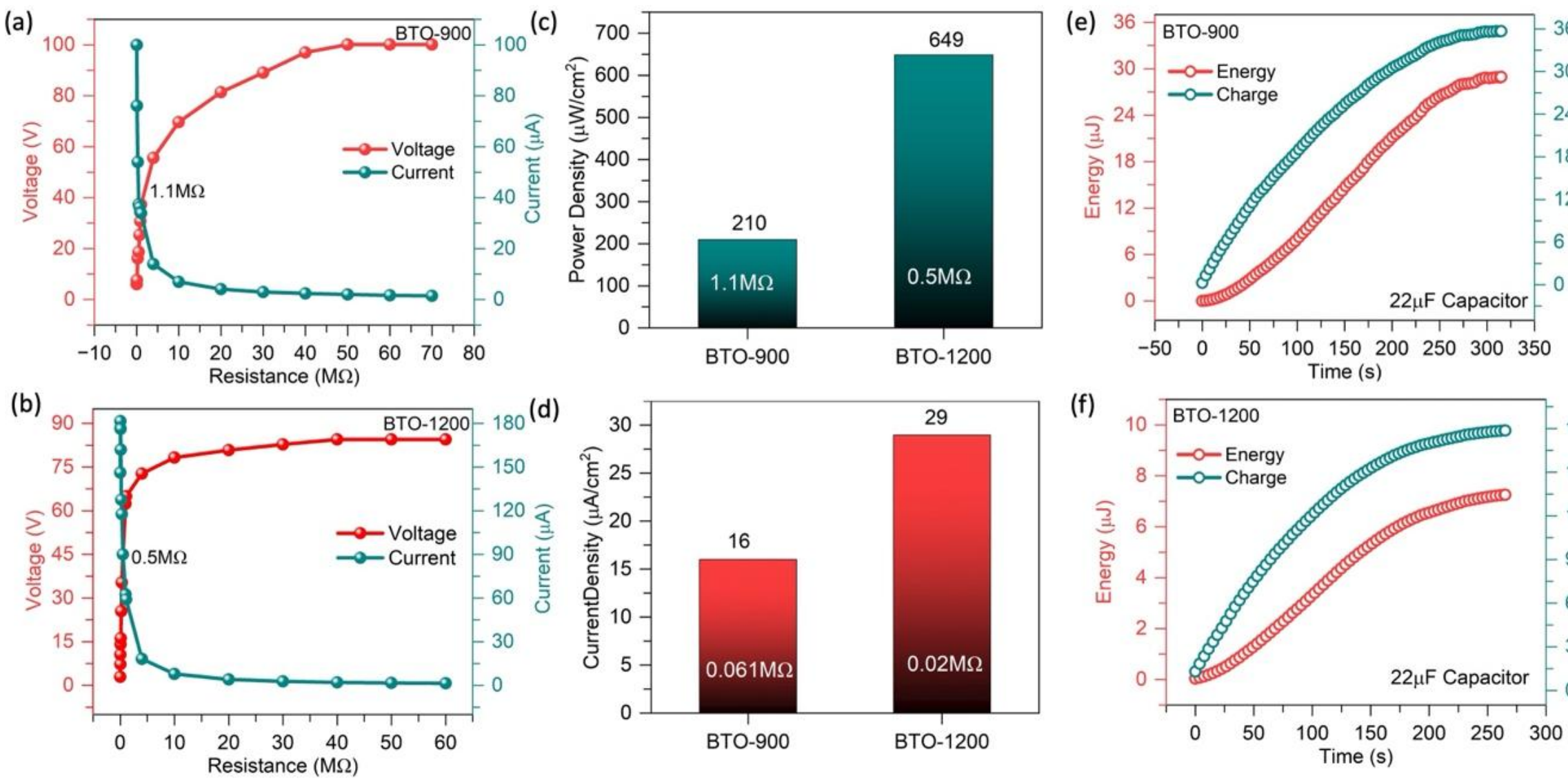

**Figure 6** (a,b) I-V-R plot showing ~100 V and ~85 V saturation Voc, Isc of ~100 μA and 181 μA; (c,d) power density and current density; (e, f) 22 μF capacitor charging for BTO-900 and BTO-1200 respectively.

detailed in experimental section, BTO NPs were dispersed into PDMS matrix at different weight % in order to optimize the nanogenerator response.

Incorporating ferroelectric BTO nanoparticles into the PDMS layer enhances the piezoelectric potential generated under cyclic compression while simultaneously raising the dielectric permittivity and charge-storage capability of the composite; the latter reinforces the triboelectric charge generated at the contact interface through improved charge trapping and interfacial polarization.[34] The dominant contribution to the electrical output arises from the piezoelectric potential generated by the ferroelectric BTO nanoparticles during cyclic compression, which is further reinforced by triboelectric charging at the interface through piezoelectric-triboelectric coupling.[35]

Figures 5 (c,d) show the output voltage response of the BTO-900 and BTO-1200 based devices at different filler concentrations. Both devices exhibit stable alternating voltage signals under periodic mechanical excitation, confirming effective piezoelectric energy harvesting behavior. The output voltage increases with increasing BTO concentration and reaches an optimum response at 15 wt% for BTO-900 and 20 wt% for BTO-1200. Beyond the optimum concentration, a reduction in output is observed, which may be attributed to particle agglomeration and reduced stress transfer efficiency within the composite matrix. The comparatively higher output generated by BTO-900 indicates its enhanced electromechanical response and polarization characteristics.

The higher Voc of BTO-900 is consistent with the DFT, PDF, and ELF analyses, which revealed enhanced Ti-O covalency, localized electron density, and a narrower HOMO-LUMO gap that support efficient charge separation and stronger dipole alignment.

**3.8 Power Density and Current Density**

The output voltage and current strongly depend on external load resistance. Figure 6 (a, b) show the measurements of voltage and current as a function of resistance for BTO-900 and BTO-1200 respectively. As the load resistance increases, the output voltage approaches Voc, while the current decreases, following the behavior of a high-impedance Ohmic source with finite internal resistance:

$$I_{SC} = \frac{V_{OC}}{R_L + R_{int}} \quad (3)$$

where $R_L$ is the load resistance and$R_{int}$ is the internal resistance of the nanogenerator [36]. Maximum power output occurs near impedance matching between external and internal resistance.[36] Accordingly, peak power density as well as current density were achieved near ~1.1 MΩ for BTO-900 and ~0.5 MΩ for BTO-1200 as shown in Figure 6 (c,d).

The enhanced performance of the BTO-PDMS composites originates from the high dielectric constant of BTO nanoparticles. The incorporation of BTO increases the effective dielectric permittivity of the PDMS matrix, improving charge storage capability and triboelectric surface charge density.[34] The BTO-polymer interface also promotes Maxwell-Wagner-Sillars (MWS) interfacial polarization, leading to charge accumulation and suppressed charge dissipation during cyclic operation.[37]
The open-circuit voltage of the triboelectric nanogenerator can be expressed as:

$$V_{OC} = \frac{\sigma\, x(t)}{\varepsilon_0 \varepsilon_r} d \tag{4}$$

where $\sigma$is the surface charge density, $x(t)$is the separation distance, $\varepsilon_0$is vacuum permittivity, $\varepsilon_r$is the dielectric constant, and $d$is the dielectric thickness.[36]
The Voc of piezoelectric nanogenerator device is given by [33],

$$V_{OC} = \frac{d_{33}\, F\, d}{\varepsilon_0 \varepsilon_r\, A} \tag{5}$$

where, $d_{33}$ is piezoelectric coefficient, $F$ is applied mechanical force, $d$ is thickness of film, $A$ is effective area of film.
The internal capacitance of the triboelectric layer is given by:

$$\mathrm{C} = \frac{\varepsilon_0 \varepsilon_r\, A}{d} \tag{6}$$

where $A$ is the effective contact area.[36] Higher dielectric permittivity therefore enhances capacitance and improves device performance.

The relatively low internal impedance of the BTO-PDMS composites indicates efficient charge transport and interfacial coupling. Although the output voltage is lower than some porous contact-separation triboelectric systems, the achieved short-circuit current (~181 µA) is comparable to or higher than many previously reported BTO-based piezoelectric and hybrid nanogenerators containing graphene quantum dots or aligned nanofibers.[38,39]
The contrasting performance of BTO-900 and BTO-1200 arises from differences in particle size, defect concentration, and charge transport behavior. BTO-900 (~54 nm) due to higher surface-to-volume ratio possesses a greater concentration of oxygen-vacancy-related defects, which act as charge-trapping centers and improve charge retention.[29] Consequently, BTO-900 exhibits higher internal impedance (~1.1 MΩ) and superior capacitor-charging capability. The stored energy in the capacitor was calculated using:

$$E = \frac{1}{2} CV^2 \tag{7}$$

The prolonged current pulse duration in BTO-900 increases the total transferred charge, enabling a 22 µF capacitor to store ~29 µJ energy and reach ~1.63 V after 300 s (Figure 6(e)).
On the other hand, calcination at 1200 °C promotes grain growth (~104 nm) and reduces oxygen-vacancy related trap states.[29] The lower internal impedance (~0.5 MΩ) facilitates rapid carrier transport and generates a larger instantaneous short-circuit current (~181 µA) with a higher matched-load power density (~649 µW $cm^{-2}$). However, rapid charge relaxation between successive compression cycles causes partial charge dissipation before transfer to the external capacitor, limiting the stored energy to ~7.5 µJ and the capacitor voltage to ~0.83 V (Figure 6(f)).
It is noteworthy that higher instantaneous power density does not necessarily correspond to superior capacitor-charging performance in such nanogenerator systems. Instead, effective energy storage is governed by cumulative charge transfer, interfacial trapping, and impedance-controlled charge retention within the BTO-PDMS nanocomposites. **Table-2** shows comparison between different reported BTO-based piezoelectric and triboelectric nanogenerator devices with current work.

**Table 2** Comparison of reported BTO-based piezoelectric and triboelectric nanogenerator devices with the current work

| Composition | Mode of Operation | ($V_{OC}$) | ($I_{SC}$) (A) / (J) ($Acm^{-2}$) | Power /Power Density (PD) | Impedance | Ref. |
|---|---|---|---|---|---|---|
| 7wt% BTO microparticles + porous PDMS | TENG | 2500 V | 150 µA | 120 µW$cm^{-2}$ | 5 MΩ | [34] |
| Poled BTO NP + PDMS | TENG | 375 V | 6 µA | 2.25mW | 100 MΩ | [40] |
| BTO NP + Graphene Quantum Dots (GQD) + PDMS | TENG | 310 V | 23 µA | 160 µW$cm^{-2}$ | 5 MΩ | [38] |
| BTO Nanowires + PDMS | TENG | 7 V | 0.36 µA | 1.2 µW | 20 MΩ | [41] |
| Aligned BTO Nanofibers + PDMS | TENG | 2.67 V | 0.26 µA | 0.18 µW | 10 MΩ | [39] |
| 23vol% 300nm BTO NP + PDMS | TENG | 10 V | 2.17 µA | NA | NA | [42] |
| BTO + PVDF Nanofibers | PENG | 50 V | 0.0312 µA$cm^{-2}$ | 0.407 µW$cm^{-2}$ | 10 MΩ | [43] |
| BTO Microparticles + PDMS | TENG | 4 V | NA | 0.035 µW$cm^{-2}$ | NA | [44] |
| Poled BTO NP + PDMS | PENG+TENG | 13.5 V | 17.2 $\mu Acm^{-2}$ | NA | 10 MΩ | [45] |
| 10vol% $SrTiO_3$ NP + PDMS | TENG | 338 V | 9.06 $\mu Acm^{-2}$ | 647 $\mu Wcm^{-2}$ | 11 MΩ | [6] |
| $BaTi_{(1-x)} ZrxO_3$ + PVDF | PENG | 11.9 V | 1.35 μA | 15.8 nW | NA | [46] |
| 0.3vol% $Ti_{0.8}O_2$ nanosheets + 1.5vol% Ag NP+ BaTiO3 NP | PENG+TENG | 150 V | 0.32 $\mu Acm^{-2}$ | 320 $\mu Wcm^{-2}$ | 80 MΩ | [47] |
| **20% BTO-1200 NP + PDMS** | PENG+TENG | 85 V | 181.60 μA, 28.96 $\mu Acm^{-2}$ | 649 $\mu Wcm^{-2}$ | 500 kΩ | **This Work** |
| **15% BTO-900 NP + PDMS** | PENG+TENG | 100 V | 100 μA, 16 $\mu Acm^{-2}$ | 210 $\mu Wcm^{-2}$ | 1.1 MΩ | **This Work** |

## 4. Conclusions

In summary, tetragonal $BaTiO_3$ nanoparticles with controlled structural distortion were successfully synthesized through a ligand-assisted sol-gel route and systematically investigated as the active piezoelectric phase of hybrid piezoelectric nanogenerators (PENGs) operating with triboelectric synergy. Structural analyses confirmed pseudo-cubic tetragonal BTO-900 and well-defined tetragonal BTO-1200, while DFT calculations revealed enhanced Ti-O covalency, stronger Ti 3d-O 2p hybridization, narrower HOMO-LUMO gap, and improved dipole coherence in BTO-900. These electronic features promote stronger ferroelectric polarization and efficient charge retention.

Although BTO-1200 exhibited a higher instantaneous power density (~649 $\mu W\ cm^{-2}$) and short-circuit current due to its lower internal impedance and faster carrier transport, BTO-900 demonstrated superior practical energy-harvesting capability. The smaller particle size (~54 nm) of BTO-900 resulted in a larger surface-to-volume ratio and higher concentration of oxygen-vacancy-related defect states, which act as charge-trapping centers and improve interfacial polarization and charge retention. Consequently, the BTO-900 device generated a higher open-circuit voltage (~100V) and exhibited significantly better capacitor-charging performance, storing ~29 μJ energy compared to ~7.5 μJ for BTO-1200. These results demonstrate that, for self-powered piezoelectric nanogenerator systems, effective energy storage is governed not only by instantaneous power density but also by impedance-controlled charge retention and cumulative charge transfer.

Importantly, this work establishes that nanoparticle size and bond dynamics provides an effective pathway to tune ferroelectric polarization, defect concentration, and interfacial charge transport in BTO-based nanocomposites. By controlling particle size and calcination conditions, the balance between charge transport and charge retention can be optimized for targeted piezoelectric nanogenerator performance. The findings therefore highlight defect-engineered nanoscale tetragonal BTO as a promising lead-free material platform for next-generation self-powered electronics, wearable sensors, and sustainable energy-harvesting systems.

**Author contributions**

Shivshankar Jokare: Methodology, Software, Data curation, Formal analysis, Validation, Investigation, Writing - original draft, Visualization, Resources. Vikash Kushwaha: Investigation, Resources, Formal analysis, Methodology, Data curation. Mandar Shirolkar: Formal analysis, Software, Validation, Visualization. Shruti Kharadkar: Methodology, Software, Data curation. R. Boomishankar: Writing - review & editing, Formal analysis, Supervision, Resources, Validation. Smita Chaturvedi: Conceptualization, Funding acquisition, Investigation, Writing - original draft, Writing - review & editing, Methodology, Validation, Visualization, Formal analysis, Software, Project administration, Resources, Supervision, Data curation.

**Author Details**

1) Shivshankar Jokare (shivbasav00@gmail.com): https://orcid.org/0009-0000-7948-4607
2) Vikash Kushwaha (vikash.kushwaha@students.iiserpune.ac.in): https://orcid.org/0000-0002-0342-9721
3) Mandar Shirolkar (mmshirolkar@gmail.com) : https://orcid.org/0000-0003-2508-1460
4) Shruti Kharadkar (kharadkarshruti276@gmail.com): https://orcid.org/0009-0003-1582-5240
5) Boomishankar Ramamoorthy (boomi@iiserpune.ac.in): https://orcid.org/0000-0002-5933-4414
6) Smita Chaturvedi (smita.chaturvedi24@gmail.com): https://orcid.org/0000-0001-7744-9605

**Conflicts of interest**

There are no conflicts to declare.

**Data availability**

All data is available from corresponding author upon request. Supplementary data is available containing XRD images and reaction mechanisms.

**Acknowledgements**

Authors acknowledges Department of Physics and Interdisciplinary School of Science, Savitribai Phule Pune University, Pune, India, Indian Institute of Science Education and Research, Pune, India for providing facilities and lab for synthesis and characterizations. Authors acknowledge Mr. Nilotpal Deka from IISER Pune, for assisting in measurements. SC acknowledges the ESRF for provision of synchrotron radiation facilities and Momentum Transfer for facilitating the measurements. SC also acknowledges DST TAR/2022/000621 from the Department of Science and Technology, Ministry of Science and Technology, India and Prof. Surjeet Singh, IISER Pune, for hosting the DST TAR/2022/000621 project at IISER Pune. R.B. thanks, the ANRF India via grant number ANRF/ARG/2025/001573/CS.

**Notes and references**